# School reopening should be guided by solid evidence and mitigation measures against Covid-19

R. Battiston (1)(2) (c.a.), F. Carinci (3), A. Ferretti (4), R. Iuppa (1)(2), G. Lattanzi (1),
D. Menasce (5), S. Merler (6), M. Mezzetto (7), R. Potestio (1), G. Salina (8), C. Spinella(9),
S. Tonini (1), L. Tubiana (1), Y. Velegrakis (1)(10), R. De Vogli (11)

\* Working group of the Observatory for Epidemiological Data, University of Trento

(1) Physics Department, University of Trento, Italy
(2) INFN-TIFPA Trento, Italy
(3) Department of Statistical Sciences, University of Bologna, Italy
(4) Physics Department, University of Turin and INFN – Section of Turin , Italy
(5) INFN, Section of Milano Bicocca and COVIDSTAT, Italy
(6) Fondazione Bruno Kessler, Trento, Italy
(7) INFN, Sezione di Padova and INFN-COVIDSTAT, Italy
(8) INFN, Section of Roma Tor Vergata, Italy
(9) CNR IMM, Catania, Italy
(10) Data Intensive Systems Group, Utrecht University
(11) Department of Social Psychology and Development, University of Padova

The debate on the role of school closures as a mitigation strategy against the spread of Covid-19 is gaining relevance due to emerging variants in Europe. According to WHO, decisions on schools "*should be guided by a risk-based approach*"[1]. However, risk evaluation requires sound methods, transparent data and careful consideration of the context at the local level[2-9].

Such fundamental criteria are at odds with a recent study by Gandini et al., on the role of school opening as a driver of the second COVID-19 wave in Italy[10], which concluded that there was no connection between school openings/closures and SARS-CoV-2 incidence. infections. This analysis has been widely commented in Italian media as conclusive proof that "*schools are safe*"[11]. However the study presents severe oversights and careless interpretation of data.

The authors claim that school closures did not affect the R(t) trend in italian regions Lombardy and Campania. Data presented in Figures 5A&B are however flawed by temporal misalignment errors: R(t) presented in Figure 5 refers to later time periods, 6 days for Lombardia and 21 days for Campania. Moreover, school opening and closing in Campania took place on Sep. 24th, and Oct. 16th, not Sep. 14th and Oct. 26th as marked in the figure.

As we show in our Figure 1a[12], these errors undermine the conclusions of the study; after correct alignment, R(t) values indeed lend support to the hypothesis that school reopenings and closures are followed by increases and reductions of R(t), respectively.

In Figure 1 (right)[10], authors show a "*2-fold higher*" incidence of SARS-CoV-2 among teachers and school staff compared to the general population. They claim that this difference disappears when compared only to the 25-65 age group: however they only analyze the Veneto region and only until Oct. 17th.

As we show in our Figure 2a[12], when the incidence among teachers is compared to the working age population at the national level and on the full interval until Nov. 7th, the two-fold incidence does not disappear.

The authors seem to downplay evidence showing that the incidence of SARS-CoV-2 among students aged 14-18 years is only slightly lower than that of the general population. However, the incidence among students under 18 cannot be compared to the incidence of the general population without considering the higher fraction of asymptomatic cases w.r.t the adult population, which can lead to severe underestimation of positive cases in younger age groups, also considering the extremely variable use of test swabs among schools[12].

By analyzing the effects of school openings the authors state that "the *increase in R(t) in different Italian regions occurred indeed after school opening*" but they could not find "*an unequivocally constant delay between school opening and R(t) rise*". This latter fact should not be surprising, since they seem to assume a single, direct cause-effect relationship between these two variables. Time dependent analysis cannot be limited to the discussion of the curve shape, ignoring the complexity of multiple causal factors. Moreover, authors assume that, with no changes of government interventions, R(t) should reach zero linearly, which is an arbitrary and unproven assumption.

The authors analyze separate disjoint datasets from different populations, a practice that cannot be used to reach convincing conclusions. Open data are not always used along the paper, while they represent a crucial point to ensure quality by design.

This article suggests the viability of keeping schools open while reporting the contradicting evidence of a double incidence of Covid-19 among members of the school staff. After correcting the misalignments of the paper, school openings actually precede rising Rt values; schools closing, on the contrary, happens before Rt reduction. These results are in stark contrast with the conclusions of the authors.

Any investigation addressing critical decisions, e.g. school closures and openings, should be based on properly designed studies. By no means lack of evidence can be interpreted as evidence of absence.

Teachers, school staff and to a lesser extent 13-18 years old students may indeed be at risk of being infected by SARS-CoV-2 and infect others. As suggested by a recent Lancet editorial, schools can be made really safe if a series of strategies including frequent testing, contact tracing and effective isolation, is implemented.

**References**

1. WHO International. Coronavirus disease: schools. Available at: *https://www.who.int/emergencies/diseases/novel-coronavirus-2019/question-and-answers-hub/q-a-detail/coronavirus-disease-covid-19-schools* Accessed on April 2$^{nd}$ 2021.
2. The Guardian. France to close schools and stop domestic travel after Covid surge. Wed 31st Mar 2021. Available at: https://www.theguardian.com/world/2021/mar/31/macron-to-unveil-tough-measures-as-covid-cases-surge-in-france Accessed 2nd April 2021.
3. Naimark D, Mishra S, Barrett K, Khan YA, Mac S, Ximenes R, Sander B. Simulation-Based Estimation of SARS-CoV-2 Infections Associated With School Closures and Community-Based Nonpharmaceutical Interventions in Ontario, Canada, JAMA Netw Open. 2021 Mar 1;4(3):e213793. Available at: https://jamanetwork.com/journals/jamanetworkopen/fullarticle/2777976 Accessed 2nd April 2021.

4. Liu Y, Morgenstern C, Kelly J, Lowe R, Jit M. The impact of non-pharmaceutical interventions on SARS-CoV-2 transmission across 130 countries and territories, BMC Med. 2021 Feb 5;19(1):40. Available at: https://www.ncbi.nlm.nih.gov/pmc/articles/PMC7861967/  Accessed 2nd April 2021.
5. Staguhn ED, Weston-Farber E, Castillo RC. The impact of statewide school closures on COVID-19 infection rates. Am J Infect Control. 2021 Apr;49(4):503-505. Available at: https://www.ncbi.nlm.nih.gov/pmc/articles/PMC7831551/  Accessed 2nd April 2021.
6. Brauner JM, Mindermann S, Sharma M, Johnston D, Salvatier J, Gavenčiak T, Stephenson AB, Leech G, Altman G, Mikulik V, Norman AJ, Monrad JT, Besiroglu T, Ge H, Hartwick MA, Teh YW, Chindelevitch L, Gal Y, Kulveit J. Inferring the effectiveness of government interventions against COVID-19. Science. 2021 Feb 19;371(6531):eabd9338.  Available at: https://www.ncbi.nlm.nih.gov/pmc/articles/PMC7877495/  Accessed 2nd April 2021
7. Ulyte A, Radtke T, Abela IA, Haile SR, Berger C, Huber M, Schanz M, Schwarzmueller M, Trkola A, Fehr J, Puhan MA, Kriemler S. Clustering and longitudinal change in SARS-CoV-2 seroprevalence in school children in the canton of Zurich, Switzerland: prospective cohort study of 55 schools. BMJ. 2021 Mar 17;372:n616.  Available at: https://www.ncbi.nlm.nih.gov/pmc/articles/PMC7966948/ Accessed 2nd April 2021.
8. Buonsenso D, De Rose C, Moroni R, Valentini P. SARS-CoV-2 Infections in Italian Schools: Preliminary Findings After 1 Month of School Opening During the Second Wave of the Pandemic. Front Pediatr. 2021 Jan 14;8:615894.  Available at: https://www.ncbi.nlm.nih.gov/pmc/articles/PMC7841339/  Accessed 2nd April 2021.
9. Vlachos J,  Hertegård E, and Svaleryd HB. The effects of school closures on SARS-CoV-2 among parents and teachers. PNAS March 2, 2021 118 (9) e2020834118.  Available at: https://www.pnas.org/content/118/9/e2020834118 Accessed 2nd April 2021.
10. Gandini S, Rainisio M, Iannuzzo ML, Bellerba F, Cecconi F, Scorrano L. A cross-sectional and prospective cohort study of the role of schools in the SARS-CoV-2 second wave in Italy. Lancet Regional Health, 5, 100092, Jun 01, 2021. Available at: *https://www.thelancet.com/journals/lanepe/article/PIIS2666-7762(21)00069-7/fulltext* Accessed 2nd April 2021.
11. «*La scuola appare un luogo sicuro, non di contagio*», Corriere della Sera,  December 20[th], 2021, *https://www.corriere.it/scuola/medie/20_dicembre_19/scuola-luogo-sicuro-cosa-dicono-dati-positivi-rt-69b78032-422a-11eb-a986-08f3985f4b5a.shtml?refresh_ce-cp*
12. Repository for Figures 1a and 2a of this Commentary: https://www.dropbox.com/sh/8xi93lyk15op46v/AAB3RpOGSj5p_a-V-KA8XKRQa?dl=0

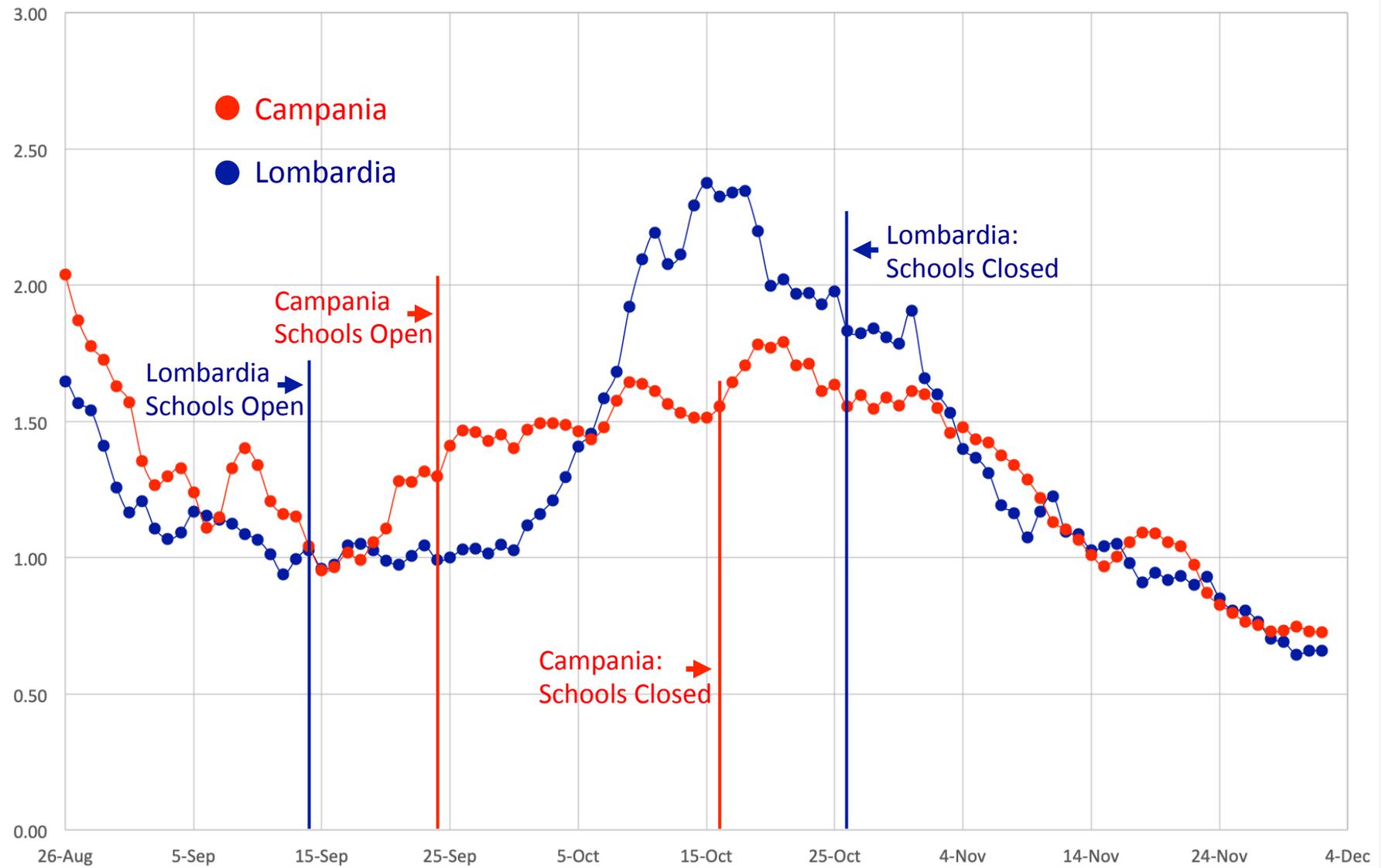

Figure 1a: Same as Figure 5A & 5B but with correct data about school opening and closing (see text of the Commentary)

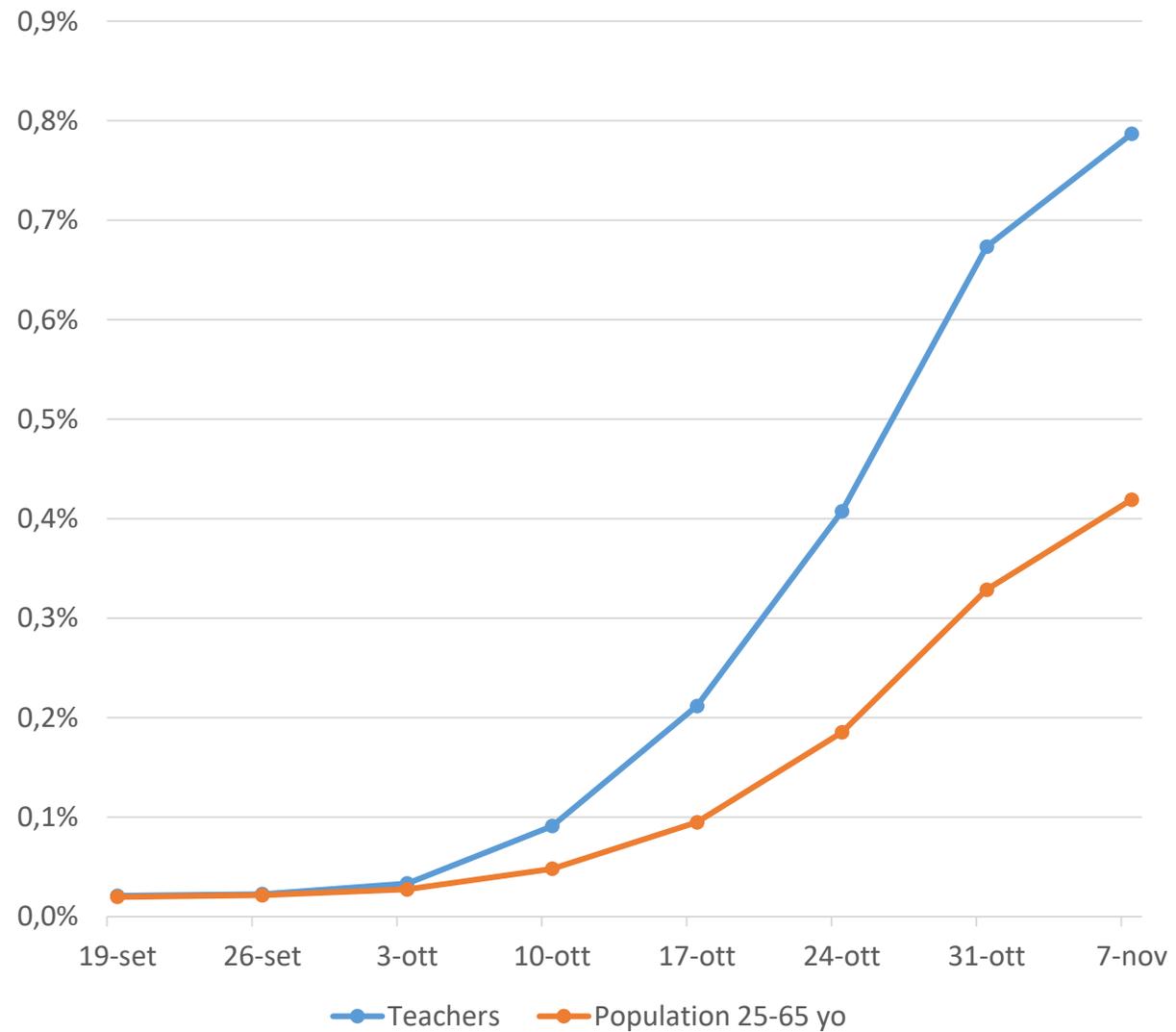

Figure 2a : a) incidence among teacher w.r.t. population 25-65 y.o.

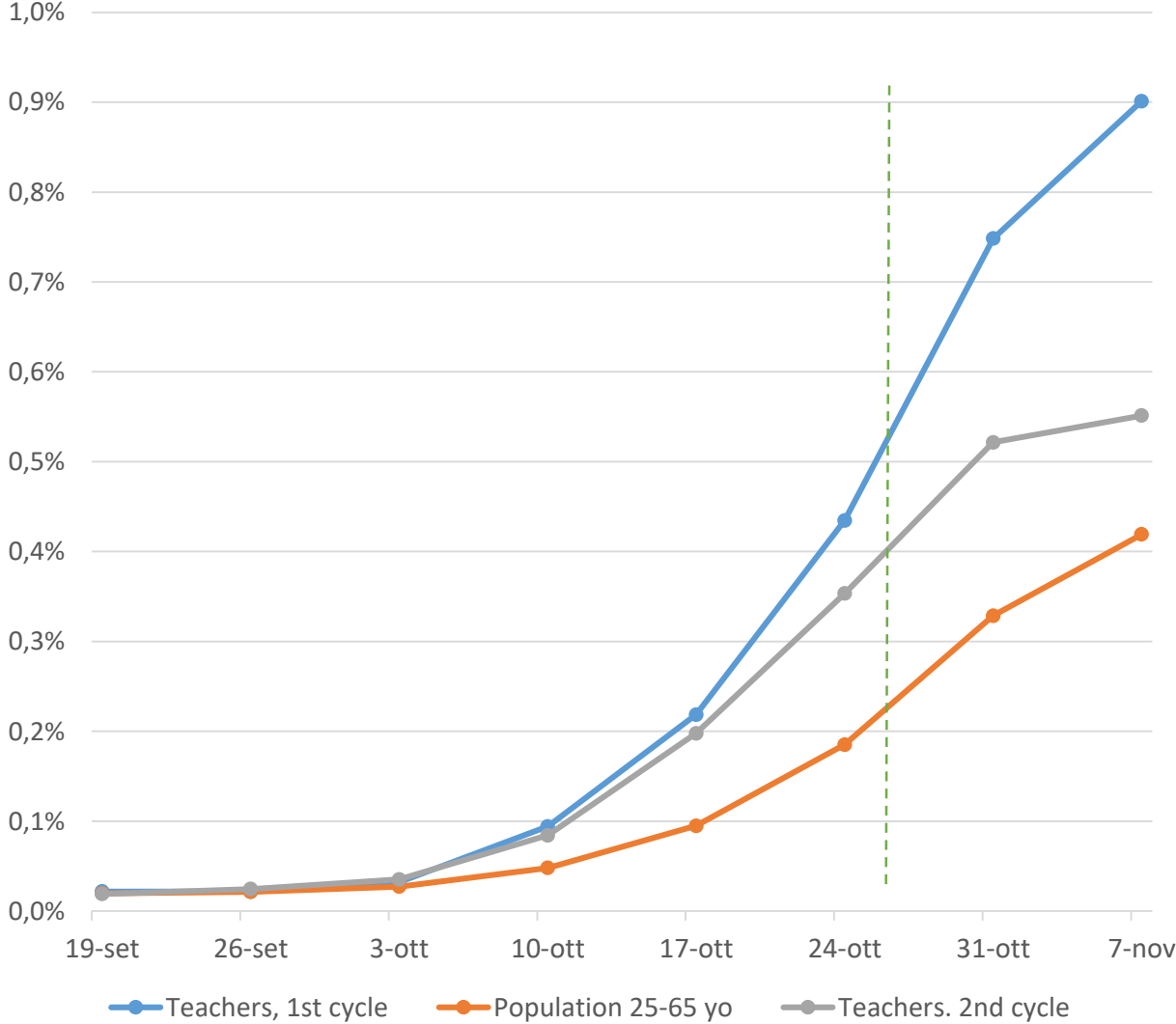

Figure 2a (b) Same as a) but teachers are split in two categories, 1st and 2nd cycle.